\documentclass[12pt]{article}

\usepackage{amsmath,amssymb}
\usepackage{graphicx,color}

\def\cH{{\mathcal{H}}}

\def\cK{{\mathcal{K}}}

\def\cW{{\mathcal{W}}}
\def\cR{{\mathcal{R}}}
\def\cL{{\mathcal{L}}}
\def\cC{{\mathcal{C}}}
\def\bH{{\bf{H}}}

\def\cM{{\mathcal{M}}}

\def\be{\beta}

\def\tg{{\tilde{g}}}

\def\tmu{{\tilde{\mu}}}

\def\r2{{\sqrt{2}}}
\def\h{{\eta}}

\def\h0{\hat{h}}
\def\Vr0{\hat{V}_{r}}
\def\Vp0{\hat{V}_{\phi}}
\def\cL{{\mathcal{L}}}
\def\r2{\sqrt{2}}
\def\be{\begin{equation}}
\def\ee{\end{equation}}

\def\bB{{\bf{B}}}
\def\bK{{\bf{K}}}

\newcommand{\Letter}{
\setlength{\textwidth}{16.5cm}
   \setlength{\textheight}{22.6cm}
    \hoffset=-0.6in
\voffset=-2.1cm }

\Letter

\begin{document}
\begin{titlepage}
\begin{flushright}
MAD-TH-09-02 \\
UCB-PTH-09-15
\end{flushright}
\begin{centering}
\vspace{.3in}

{\Large {\bf On D3-brane Dynamics at Strong Warping}}


\vspace{.3in}

Heng-Yu Chen${}^1$, Yu Nakayama${}^2$, and Gary Shiu${}^{1,3}$ \\
\vspace{.2 in}
${}^{1}$\textit{Department of Physics, University of Wisconsin,
Madison, WI 53706, USA}

\vskip 3pt
\vspace{.1 in}
${}^{2}$\textit{Berkeley Center for Theoretical Physics and Department of Physics, \\
University of California, Berkeley, CA 94720, USA}

\vskip 3pt
\vspace{.1 in}
${}^{3}$\textit{School of Natural Sciences, Institute for Advanced Study, Princeton, NJ 08540, USA}

\bigskip \bigskip \bigskip

{\bf Abstract}

\vspace{.1in}
We study the dynamics of a D3 brane in generic IIB warped compactifications, using the Hamiltonian formulation discussed in \cite{DT}. 
Taking into account of both closed and open string fluctuations, we derive the warped K\"ahler potential governing the motion of a probe D3 brane.
By including the backreaction of D3, we also comment on how the problem of defining a holomorphic gauge coupling on wrapped D7 branes in warped background can be resolved.    
\end{centering}

\vfill

\begin{flushleft}
\today
\end{flushleft}

\end{titlepage}

\section{Introduction}
\paragraph{}
A recurrent theme in recent developments of 
string theory has been the utility of  warped geometries 
in an increasingly wide variety of physics contexts. In additional to its pivotal role in understanding strongly coupled field theories via the
 gauge/gravity correspondence \cite{Maldacena:1997re,Gubser:1998bc,Witten:1998qj}, warping has also become an indispensable tool 
in many constructions of particle physics and cosmology models from string theory. 
The gravitational redshift due to warping 
suggests a mechanism to 
generate a hierarchy of scales\footnote{See  \cite{Verlinde,Sethi,Shiu,GKP} for some early attempts of realizing \cite{RS}, albeit with supersymmetry, in string theory.}, thus realizing
in a geometric way \cite{RS} the idea of technicolor.
Even  within the framework of supersymmetry, 
the omnipresent
strongly coupled hidden sectors 
often
admit gravity duals
that are 
warped
and as such provide a
 holographic description of supersymmetry breaking and its mediation (see, e.g., \cite{Nakayama:2007cf,Benini:2009ff}).
Likewise,
warping
enables
us to determine the structure of the 
inflation potential for some string inflationary scenarios \cite{stringcosmologyreviews}
by 
utilitizing 
the 
systematics of 
the 
AdS/CFT 
duality
\cite{Baumann:2008kq},
not to mention that it also introduces
tunable small numbers in the inflaton potential
useful for  model building purposes.

Given the variety of applications, it is of interest to determine the low energy effective action
governing string theory on strongly warped backgrounds.
In particular, to draw precise predictions in models of particle physics and cosmology constructed in this framework, one needs to take into account all warping contributions to the effective action, including  their effects on the K\"ahler potential
which is not protected by holomorphy\footnote{The inflaton potential and the flavor problem in gravity mediation are two examples in which the precise form of the K\"ahler potential plays an important role.}.
While the 
vacuum structure 
of warped compactifications have been well explored, 
issues involving the {\it dynamics} of such backgrounds are much less understood. 
In fact, the derivation of 
a warped effective theory describing fluctuations around
a strongly warped background
has proven to be highly subtle \cite{GM,Burgess:2006mn,STUD,Martucci:2009sf}.
Though significant progress has recently been made in
computing the effective action
for the closed string sector \cite{DT,STUD} (and in some cases,
simple expressions
of the warped K\"ahler potential
were presented  \cite{FTUD}),
the inclusion of open string degrees of freedom 
has so far  been carried out \cite{MMS} (see also \cite{Martucci:2006ij}) only in the D-brane probe limit and in a fixed closed string background.
By demanding 
the
combined 
K\"ahler potential
to reproduce 
the correct kinetic terms for the open string fluctuations in a warped background,
one can constrain its form \cite{MMS}, and indeed the results 
find agreement with previous closed string computations.  
However, to go beyond these limited examples,  
it is important to 
extend the analysis of \cite{MMS} to allow for both open and closed string fluctuations.
For some applications of warped compactifications such as D-brane inflation where both open and closed string degrees of freedom are simultaneously dynamical, this extension is in fact inevitable.

The aim of this note is to take a first step in this direction by considering a D3-brane in a strongly warped background.
We include the fluctuations of both the D3 moduli and the closed string moduli in our analysis and derive the combined K\"ahler potential.
Among the subtleties in deriving the warped effective action is a correct identification of the gauge invariant perturbations \cite{DT,STUD}.
A naive dimensional reduction without
dividing the field space metric by
gauge redundancies 
leads to a conjectured expression \cite{DG}
that does not minimize the inner product of metric perturbations over each gauge orbit.
We identified the gauge invariant perturbations for the combined open and closed modul space and obtained the resulting K\"ahler potential.
Although our main focus is the combined open and closed string system, our results have clarified and elucidated some previous closed string computations. In particular,
we found a gauge that simplifies not only  the worldvolume action of the 
D3-branes at strong warping, but also in computing the K\"ahler metric for the
closed string fluctuations.

This paper is organized as follows. In Section \ref{HamiltonianFormalism}, we review and clarify the Hamiltonian formalism which we used to extract the kinetic terms of moduli fields in warped compactification.
In particular, we discuss how one can carry out the computation in different gauges, emphasizing that 
the K\"ahler metrics are gauge invariant while the Hamiltonian constraints (except for special cases  discussed below) depend on the gauge choice.
We also suggest a convenient
gauge useful for deriving the kinetic terms of light scalar fields 
that do not develop a classical potential\footnote{This does not exclude the possibility that those moduli are lifted by non-perturbative effects, e.g., the universal K\"ahler modulus in \cite{KKLT}.}.
In Section \ref{CouplingtoD3}, we extend our approach to include both 
fluctuating closed string and D3-brane moduli.
The convenient gauge choice introduced in Section \ref{HamiltonianFormalism} also turns out to diagonalize the kinetic terms for the  open plus closed string system.
As an illustration, we present the combined K\"ahler potential involving the D3 moduli and the universal K\"ahler modulus.
We also comment on how the rho problem can be solved even in the strong warping limit when backreaction of the D3-branes is taken into account. We end with some discussions in Section \ref{Discussions}.

\section{Warped Compactification and Hamiltonian Formalism}\label{HamiltonianFormalism}
\paragraph{}
We first review the relevant background about extracting the kinetic terms of moduli fields in warped compactification, following \cite{DT,GM,STUD}.
Our starting point is the supergravity action for the type IIB string theory in Einstein frame. We follow the convention in \cite{GKP}:
{\small \begin{equation}
S_{IIB}^{\rm E}=\frac{1}{2\kappa_{10}^2}\int d^{10}x\sqrt{-g_{10}}\left\{\cR^{(10)}-\frac{\partial_M\tau\partial^M\tau}{2({\rm Im}\tau)^2}-\frac{G_3\cdot\bar{G}_3}{12{\rm Im}\tau}-\frac{\tilde{F}_5^2}{4 \cdot 5!}\right\}
+\frac{1}{8i\kappa_{10}^2}\int\frac{C_4\wedge G_3\wedge\bar{G}_3}{{\rm Im}\tau}+S_{\rm loc.}\,.\label{IIBaction}
\end{equation}}
Here $\tau$ is the complex dilaton-axion, $G_3=F_3-\tau H_3$ is the complex three form flux, $C_4$ is the RR four form and $\tilde{F}_5=dC_4-\frac{1}{2}C_2\wedge H_3+\frac{1}{2}B_2\wedge F_3$ is the self-dual five form flux, satisfying $\tilde{F}_5=\star_{10} \tilde{F}_5$.
The term $S_{\rm loc.}$ is the action of localized objects such as D-branes, whose kinematics in a warped background will be studied momentarily.

We will focus on the ten dimensional warped metric which preserves the maximal four dimensional isometries. It takes the form:
\begin{equation}
ds^2=e^{2A(y,u)}\tg_{\mu\nu}(x)dx^\mu dx^\nu+e^{-2A(y,u)}\tg_{ij}(y,u)dy^i dy^j\,.\label{staticmetric}
\end{equation}
Here $e^{-4A(y,u)}$ is the warped factor and its precise form depends on the fluxes and localized sources. For definiteness, we will consider the supersymmetric backgrounds generated by the fluxes satifying \cite{GKP}:
\begin{equation}
e^{4A_0}=\alpha(y)~~{\rm where}~~ \tilde{F}_5=(1+\star_{10})[d\alpha(y)\wedge d^4 x]\,,\label{backe4A}
\end{equation}
and the Imaginary Self-Dual (ISD) condition
\begin{equation}
\star_6 G_3= iG_3\,.\label{ISD}
\end{equation}
Given the background solutions, the whole set of zero modes $\{u^I\}$, such as the complex and K\"ahler structure moduli, parametrize the fluctuations around them.  
Under the dimension reduction to the four dimensional spacetime, we can extract the kinetic energy terms for $u^I$ in the four dimensional effective field theory by promoting them into spacetime dependent fields $u^I(x)$. The metric $G_{IJ}(u)$ on the space of zero modes parameterized by $\{u^I\}$ or ``the moduli space metric'' in short, can subsequently be obtained from the terms quadratic in $\partial_\mu u^I(x)$:
\begin{equation}
\int d^4 x\sqrt{-\tilde{g}_4} \tilde{g}^{\mu\nu} G_{IJ}(u)\partial_\mu u^{I} \partial_\nu u^{J}\,.\label{4DKE}
\end{equation} 
However not all the metric fluctuation $\delta_I g_{MN}(u)$ should enter the computation of (\ref{4DKE}), as
further subtleties arise.
It is illuminating to pause here and draw parallel with the moduli space metric of Yang-Mills instantons (for a modern introduction, see e.g. \cite{InstantonReview}). Suppose $A_\mu^{(0)}$ is an instanton solution and consider a small fluctuation $\delta_I A_\mu$ around $A_\mu^{(0)}$, as parameterized by some moduli/collective coordinates $\{X^I\}$. For $\delta_I A_\mu$ to be a {\it physical} zero mode that enters the moduli space metric of instanton, it needs to satisfy both {the linearized equation of motion} and {the orthogonality condition to other gauge transformations}.

The situation with the metric zero modes is analogous. First, to ensure that the perturbed metric remains a solution to the ten-dimensional equation of motion, it is necessary for the metric fluctuations to satisfy the linearized equation of motion.
Generically this fails by naively setting $\{u^I\}$ to be dynamical in (\ref{staticmetric}), and more elaborated ansatz is required.     
The presence of fluxes and localized sources typically makes explicitly solving the linearized equation a difficult task.

Moreover for a given metric ansatz satisfying the linerized equations of motion, an infinite family of equivalent $u^I$-dependent metric zero modes can be futher generarated by the ten dimensional diffeomorphism transformation. This is analogous to the gauge transformation in Yang-Mills theory.  
As for the instantons, on the space of possible metric zero modes, the {\it physical} or {\it inequivalent} metric fluctuations which appear in the moduli space metric are the ones orthogonal to the gauge/diffeomorphism transformations. 
Additional constraint on $\delta_I g_{MN}$ is therefore required to ensure such an orthogonality condition. 

Given the linearized equations of motion and the constraint from orthogonality, the analysis for metric zero modes hence the moduli space metric becomes seemingly involved.
As it turned out, however, for extracting the modular kinetic energy terms, an elegant and efficient approach based on the familiar ADM/Hamiltonian formulation of general relativity \cite{ADM} (see also \cite{Wald}) was introduced in \cite{DT}. 
In this approach, by using a particular type of ansatz for the metric fluctuations, the linearized equations of motion containing single time derivative naturally arise as the gauge orthogonality conditions.
\paragraph{}
The starting point for applying the Hamiltonian approach with multiple moduli $\{u^I\}$ is the following general ansatz \cite{GM}:
\begin{eqnarray}
ds^2&=&e^{2A(y,u)+2\Omega(u)}
\tilde{g}_{\mu\nu}(x) dx^{\mu} dx^{\nu}+e^{-2A(y,u)}\tilde{g}_{ij}(y,u)dy^i dy^j\nonumber\\
&+&2e^{2A(y,u)+2\Omega(u)}(\partial_\mu\partial_\nu u^{I}(x) \bK_I(y)dx^\mu dx^\nu+\partial_{\mu}u^I(x) \bB_{iI}(y)dx^{\mu}dy^i)\,.\label{WarpMetric}
\end{eqnarray}
In the first line of (\ref{WarpMetric}), $u^I(x)$ are now dynamical; however for the spacetime dependent metric fluctuations, the terms in the second line, which are proportional to $\partial_\mu u^I(x)$ and $\partial_\mu\partial_\nu u^I(x)$ are also allowed. 
These additional terms are refered in the literature as the {\it compensators} \cite{GrayLukas,GM}, and they vanish when the moduli $u^I$ become static. Their presence is needed to ensure that the perturbed metric (\ref{WarpMetric}) satisfies the linearized equations of motion and they play pivotal roles in our subsequent discussions. Notice that in (\ref{WarpMetric}), the moduli are assumed to vary slowly, and, therefore, we only keep the linear order in both velocity $\partial_\mu u^I(x)$ and the acceleration $\partial_\mu\partial_\nu u^I(x)$. We will also take the magnitude of $u^I(x)$ to be small, as when $u^I$s become constant, they should correspond to small deformation of the background metric.

Under the dimensional reduction, in order to bring us back into Einstein frame where the four dimensional Planck mass is constant, here we have also included the Weyl factor $e^{2\Omega(u)}$ (see also 
\cite{FTUD}):
\begin{equation}
e^{-2\Omega(u)}=\frac{\int d^6 y \sqrt{\tg_6}e^{-4(y,u)}}{\int d^6 y\sqrt{\tg_6}}=\frac{V_W(u)}{V_{CY}}\,.\label{e2Omega}
\end{equation}
It will play an important role in our later discussion about the K\"ahler potential of various moduli. 
Strictly speaking, such a definition is only valid if the unwarped metric $\tg_{ij}$ is independent of the moduli $u^I$; otherwise the four dimensional Planck mass would be moduli dependent. For $\tg_{ij}$ with non-trivial moduli dependence, one should therefore replace $V_{CY}=\int d^6 y \sqrt{\tg_6}$ by other moduli independent fiducial volume measure, such as $(\alpha')^3$.
Notice that in constrast to earlier literatures (see for example \cite{Baumann1,KMS}), where the Weyl factor (or breathing modes) and the warp factor were treated as independent quantities in the weak warping limit, here with significant warping, they are in fact related by definition.  

As the metric fluctuations become time-dependent, a more illuminating way to understand the role of compensators is to view the metric (\ref{WarpMetric}) from the familiar ADM or Hamiltonian formulation of general relativity \cite{DT} (For a detailed review, see \cite{Wald}). 
Applying this formalism to a ten dimensional spacetime $\cM$ with metric $g_{MN}$ amounts to choosing a time coordinate $t$ and dual time-flow vector $t^M$, where one can decompose $t^M$ into its normal $n^M$ and tangential $\eta^M$ components with respect to a nine dimensional space-like hypersurface of constant time $\Sigma_t$. 
We can follow the standard procedures to define the {\it lapse function} $N$:
\begin{equation}
N=-g_{MN}t^Mn^N \,,\label{lapse}
\end{equation}
and the {\it shift vector}:
\begin{equation}
\eta^M=h^{MN}\eta_N=t^M-N n^M\,,\label{shiftvector}
\end{equation}
Here $h_{MN}$ is the pullback metric of $g_{MN}$ onto $\Sigma_t$, and they are related by $h_{MN}=g_{MN}+n_M n_N$, 
we can therefore use $(h_{MN}, N\,,\eta^M)$ to recast the ten-dimensional metric into
\begin{equation}
ds^2=-N^2 dt^2+h_{MN}\left(dx^M+\eta^M dt\right)\left(dx^N+\eta^N dt\right)\,.\label{ADMmetric}
\end{equation}

Comparing (\ref{ADMmetric}) and (\ref{WarpMetric}), we clearly see the analogy.
Up to the linear order in $\partial_\mu u^I$ and $\partial_\mu\partial_\nu u^I$, the compensator can be identified with the shift vector (\ref{shiftvector}). More concretely, consider $x^0=t$ with $g_{tt}=-g_{00} > 0$ in (\ref{WarpMetric}), the lapse function is given as $N=(g_{tt})^{1/2}$.
The shift vectors and compensators are simply identified as $\eta_M=g_{0M}\,,~M=1,\dots, 9$, or in terms of components\footnote{Here we use $\alpha,\beta=1,2,3$ to denote the three spatial directions in order to distinguish $\mu,\nu=0,1,2,3$, which include time-like direction.}:
\begin{equation}
\eta_\alpha=e^{2A(y,u)+2\Omega(u)}\partial_\alpha\partial_0 u^{I} \bK_I(y)\,,~~~\eta_i= e^{2A(y,u)+2\Omega(u)} \partial_0 u^I \bB_{iI}(y)\,,~~~\alpha=1,2,3\,,~i=4\,\dots, 9.\label{DefCompensators}
\end{equation}     
It is now also immediately clear that the physical metric fluctuation should be normal to the space-like surface $\Sigma_t$, and it is given by the extrinsic curvature $K_{MN}$:
\begin{equation}
K_{MN}=\frac{1}{2}{\mathcal{L}}_n h_{MN}=\frac{1}{2}(g^{tt})^{1/2}\left(\frac{d h_{MN}}{dt}-D_M\eta_N-D_N\eta_M\right)\,.\label{KMN}
\end{equation} 
Here the covariant derivative $D_M$ is defined with respect to the nine-dimensional metric $h_{MN}$, which is identified with $h_{MN}\equiv g_{MN}\,,~M,N\neq 0$. For the remainder of the paper therefore, $M\,,N=1,\dots, 9$ unless otherwise stated. 
Notice the similarity between (\ref{KMN}) and the physical zero mode for the Yang-Mills instanton, $\eta^M$ essentially plays the role of the parameter in the compensating gauge transformation. One can also interpret it as combining the naive variation of the metric $h_{MN}$ with respect to $t$ and the diffeomorphism transformation on $\Sigma_t$.
 
Using (\ref{KMN}), one can express the Lagrangian density/Ricci scalar in terms of $K^{MN}$ given by
\begin{equation}
\cL^{G}_{\rm kin}=\sqrt{-g_{10}}\left(R^{(9)}+K^{MN}K_{MN}-(h^{MN}K_{MN})^2\right)\,.\label{DefLG}
\end{equation} 
The Hamiltonian approach involves defining the canonical momentum $\pi^{MN}$ conjugate to $h_{MN}$:
\begin{equation}
\pi^{MN}=\frac{\partial \cL_G}{\partial \dot{h}_{MN}}=\sqrt{h}\left( K^{MN}-h^{MN}(h^{PQ}K_{PQ})\right)\,.\label{PiMN}
\end{equation}
The Hamiltonian is then obtained from the definition ${\mathcal{H}}^{G}=\pi^{MN}\dot{h}_{MN}-\cL^G_{\rm kin}$, written in terms of $\pi^{MN}$:
\begin{equation}
{\mathcal{H}}^{G}=\sqrt{-g_{10}}\left(-R^9+h^{-1}\pi^{MN}\pi_{MN}-\frac{1}{8}h^{-1}(\pi_{MN}h^{MN})^2\right)+2\pi^{MN}D_M \eta_N\,.\label{DefH}
\end{equation} 
It can be shown that in the Hamiltonian, up to a total derivative term, $\eta^N$ only appears as a non-dynamical Lagrange multiplier imposing the following constraint:
\begin{equation}
D^M(h^{-1/2}\pi_{MN})=0\, .\label{constraint}
\end{equation}
We will explain the significance of (\ref{constraint}) momentarily.
The kinetic terms from the remaining Hamiltonian (modulo the nine-dimensional Ricci scalar term $R^{(9)}$) with (\ref{constraint}) imposed are then given by \cite{DT}:
\begin{equation}
\cH_{\rm kin}^{G}=\int d^9 x\sqrt{-g_{10}}h^{-1}\left(\pi^{MN}\pi_{MN}-\frac{1}{8}\pi^2\right)=\int d^9x \sqrt{h}(g_{tt})^{1/2} h^{-1/2} \pi_{MN} K^{MN}\,.
\label{HGkin}
\end{equation}
In the second equality, we highlighted that $\cH_{\rm kin}^G$ is the gravitational analog of the $p\dot{q}$ term in the classical dynamics. 
 
To understand the significance of the constraint (\ref{constraint}), we recall that there is arbitrariness in our choice of $\Sigma_t$, hence the induced metric $h_{MN}$. If $\psi$ represents the diffeomorphism of $\Sigma_t$, $h_{MN}$ and $\psi^* h_{MN}$ should represent the same physical configuration.
In other words, our physical configuration is a space of equivalence classes of metrics $h_{MN}$ on $\Sigma_t$, where the equivalence is given by the diffeomorphism $\psi$. This is precisely what is referred in the literature as the ``superspace''. 
Consider a nine dimensional diffeomorphism transformation of $\Sigma_t$ acting on $K^{MN}$:
\begin{equation}
K^{MN}~\longrightarrow~K^{MN}-\frac{1}{2}(g^{tt})^{1/2}(D^M V^{N}+D^N V^{M})\,,\label{Ktrans}
\end{equation}
The kinetic term $\cH_{\rm kin}^{G}$ is therefore shifted by a term:
\begin{equation}
-\int d^9 x \sqrt{h} D^M V^N (\pi_{MN} h^{-1/2})=\int d^9 x \sqrt{h}\left(V^{N}D^M(h^{-1/2}\pi_{MN})-D^M(V^N h^{-1/2}\pi_{MN})\right)\,.\label{ChangeKE} 
\end{equation}
As we choose the superspace as physical configuration space, the conjugate momentum $\pi_{MN}$ must ensure that (\ref{ChangeKE}) vanishes.
Since the second term in (\ref{ChangeKE}) is merely a total derivative and $V_M$ is an arbitrary vector parameterizing the diffeomorphism transformation, the constraint (\ref{constraint}) is precisely the condition needed to ensure (\ref{ChangeKE}) vanishes. 
In other words, with respect to the inner product defined in (\ref{HGkin}), (\ref{constraint}) is the condition to guarantee the metric fluctuation to be {\it physical}, i.e. orthogonal to arbitrary diffeomorphism transformation.

One can also verify that, by substituting the given ansatz (\ref{WarpMetric}), the orthogonality condition (\ref{constraint}) is equivalent to the ten dimensional linearized Einstein equations,  
since they simply come from varying the metric component $g_{0M}$. Explicitly, we can make the following identifications:
\begin{equation}
(g^{tt})^{1/2}D_N(h^{-1/2}\pi^{NM})=\delta R^{0M}=\delta G^{0M}=0\,,~~~M=1,\dots, 9\label{LinearEOM}\,.
\end{equation} 
In the second equality we have used the fact that in the limit of static $u^I$, $g_{0M}=g_{M0}=0$ and we only keep quadratic flucutuations in the action. 
In \cite{STUD}, it was also pointed out that as $\delta G_{0M}$ Einstein equations only contain single time derivative, they act as the constraints which should be satisfied by any consistent solutions at all time.   
In the Hamiltonian formalism, such constraints are made manifest and their role  is elegantly explained. 
Calculationally, the identifications (\ref{LinearEOM}) allows us to recycle some of the calculations for the linearized Ricci tensors \cite{GM}, when expressing the constraint equation (\ref{constraint}) explicitly. This also tells us that in the presence of additional localized energy sources such as D-brane, the linearized Einstein hence the constraint equations get modified.
\paragraph{}
To apply the Hamiltonian formalism reviewed earlier to extract the moduli space metric, we consider the simpler case that $\partial_\mu u^I\equiv \delta^0_\mu \dot{u}^I(t)$. The ten dimensional kinetic term is then given by \footnote{To be more precise, in our analysis for the kinetic term, we are only keeping the quadratic fluctuations. At this order the integration measure $\sqrt{-g_{10}}$ or $\sqrt{h}$ multiplying the $\pi_{MN}\pi^{MN}$ and $\pi^2$ terms do not carry any explicit $u$ dependence. However if there are additional $u$-dependence in the overall action, arising from the nine-dimensional curvature $R^{(9)}$ or flux induced potential, the $\partial_{\mu}\partial_{\nu}u^I$ terms in $\sqrt{-g_{10}}$ can in principle also give extra contribution to the kinetic energy via integration by parts.}
\begin{equation}
S_{\rm kin}^{G}=\frac{1}{2\kappa_{10}^2}\int dt \cH_{\rm kin}
=\frac{1}{8\kappa_{10}^2}\int dt \dot{u}^I\dot{u}^J\int d^9 x\sqrt{-g_{10}}g^{tt}\delta_I h^{MN}\delta_J \pi_{MN}\label{Skin}\,.
\end{equation} 
Here the terms $\delta_I h_{MN}$ and $\delta_I \pi_{MN}$ are related to $K_{MN}$ and $\pi_{MN}$ by:
\begin{equation}
K_{MN}=\frac{1}{2}(g^{tt})^{1/2}\dot{u}^I\delta_I h_{MN}\,,~~~\pi_{MN}=\frac{1}{2}\sqrt{h} (g^{tt})^{1/2} \dot{u}^{I}\delta_I\pi_{MN}\,,\label{defdelKh}
\end{equation} 
where we have factored out the moduli dependence such that $d h_{MN}/d t=\dot{u}^I(\partial h_{MN}/\partial u^I)$ and $\eta_M=\dot{u}^I\eta_{IM}$.
The moduli space metric, denoted $G_{IJ}(u)$ is then given by:
\begin{equation}
G_{IJ}(u)=\frac{1}{8\kappa_{10}^2}\int d^6 y \sqrt{\tg_6} e^{-4A+2\Omega}\delta_I \pi_{MN}\delta_J h^{MN}\,,\label{defGIJ}
\end{equation}
such that the kinetic term can be rewritten as:
\begin{equation}
S_{\rm kin}^{G}=\int d^4x \sqrt{-\tilde{g}_4} \tg^{tt} \dot{u}^I \dot{u}^J G_{IJ}(u).\label{Skin2}
\end{equation}
The constraint equation (\ref{constraint}) can also be written in terms of $\delta_I \pi_{MN}$:
\begin{equation}
D^{M}((g^{tt})^{1/2}\delta_I \pi_{MN})=0\,.\label{constraint2}
\end{equation}  

\paragraph{}
As an illustration to the earlier general discussion on the Hamiltonian formalism, let us pause here to discuss the explicit metric gauge choices where the computations will be performed. 
The presence of the compensators $\bK_I(y)$ and $\bB_{iI}(y)$ are associated with the spacetime dependent metric fluctuations. As pointed out in \cite{GM}, we can use the freedom to parametrize the flucutuations to eliminate them by performing a ten dimensional diffeomorphism transformation with the sacrifice that the internal metric and warp factor should be simultaneously changed.    
As an example, we can consider for $u^{I}(t)$, the following diffeomorphism generated by $\zeta_A$ on the metric (\ref{WarpMetric}):
\begin{equation}
{\rm K-gauge:}~~~\zeta_{\mu}=0\,,~~~\zeta_i=-e^{2A(y,u)+2\Omega(u)}\bB_{iI}(y) u^{I}(t)\,. \label{Ktrans1}
\end{equation}
Here the transformation is given by the covariant derivative defined from the full ten-dimensional metric (\ref{WarpMetric}). 
The metric (\ref{WarpMetric}) is then transformed into:
\begin{eqnarray}
ds^2_{(K)}&=&e^{2A_K(y,u)+2\Omega_{K}(u)}
\tilde{g}_{\mu\nu}(x) dx^{\mu} dx^{\nu}+e^{-2A_K(y,u)}\tilde{g}^{(K)}_{ij}(y,u)dy^i dy^j\nonumber\\
&+& 2e^{2A_K(y,u)+2\Omega_K(u)}(\ddot{u}^{I}\bK_I(y)(dt)^2)\,.
\label{WarpMetricK}\\
e^{\pm 2A_K(y,u)}&=& e^{\pm 2A(y,u)}(1\pm 2\zeta_i\partial^i A)\,,\nonumber\\
\tg_{ij}^{(K)}(y,u)&=&\tg_{ij}(y,u)-\tilde{\nabla}_i(e^{2A(y,u)}\zeta_j)-\tilde{\nabla_j}(e^{2A(y,u)}\zeta_i)\,.
\label{Kmetrics}
\end{eqnarray}
where $\tilde{\nabla}_i$ is the covariant derivative with respect to the unwarped six dimensional metric $\tg_{ij}(y,u)$.  
Notice that the transformation (\ref{Ktrans1}) shuffles the $\bB_{iI}(y)$ dependent off-diagonal terms into both the internal metric and the warp factor. In such a gauge where $\tg_{ij}^{(K)}(y,u)$ contains explicit moduli-depedence, one should replace the unwarped volume $V_{CY}$ in (\ref{e2Omega}) by a moduli-independent fiducial volume to define the Weyl factor $e^{-2\Omega_{K}}$, and define the warped volume with respect to $e^{-4A_K}$ and $\tg_{ij}^{(K)}$.    
For later convenience, we will refer to the metric ansatz (\ref{WarpMetricK}) as the {\it ``K-gauge''}. {Note that such a gauge choice is still not unique since one could perform the gauge transformation $\zeta_i = ue^{2A+2\Omega} \partial_i \Lambda $, $\zeta_\mu = - \partial_\mu u e^{2A+2\Omega}\Lambda $ to further shuffle the compensator term $\mathbf{K}$, into warp factor $e^{4A}$ as well as the internal metric $\tilde{g}_{ij}$. In the later application, we use this residual gauge freedom to define the universal K\"ahler moduli in the conventional way.

In contrast, the gauge transformation $\zeta_i =0$, $\zeta_\mu = \partial_\mu u e^{2A+2\Omega}\Lambda$ does not generate changes in the internal metric as well as the warp factor while it shuffles $\mathbf{B}_i$ and $\mathbf{K}$ \cite{GM,FTUD}. This always enables us to go from K-gauge to the gauge where $\mathbf{K} \to 0$, $\mathbf{B}_i \to \partial_i \mathbf{K}$ without changing $\tilde{g}_{ij}$ nor $e^{4A}$. This is consistent with the fact that the linearlized equations of motion \cite{GM} (or constraint equations in the Hamiltonian formulation) always accompany the combination $\mathbf{B}_{i} - \partial_i \mathbf{K}$. Note, however, one could not conversely gauge away arbitrary $\mathbf{B}_i$ in this way because $\mathbf{B}_i$ should be a total derivative to do so.


From the perspective of the Hamiltonian formulation, one can calculate the extrinsic curvature $K_{MN}$ (\ref{KMN}) in the general case (\ref{WarpMetric}) and in the K-gauge (\ref{WarpMetricK}), and demonstrate that the ten dimensional diffeomorphism transformations relating them can be decomposed into a nine dimensional diffeomorphism transformation acting on $\delta_I h_{MN}$ and a reparametrization of $g^{tt}$. 
Using the constraint (\ref{constraint2}), one can show that, with the time reparametrization in $\int dt\sqrt{\tg_{tt}} \tg^{tt} \dot{u}^I\dot{u}^J$ of the kinetic term $S_{\rm kin}^{G}$ (\ref{Skin2}) which can be ensured when one imposes time reparametrization invariance, the moduli space metric $G_{IJ}(u)$ is indeed invariant under different gauge choices.
In practical terms, if one has a metric ansatz which consistently solves the constraint equations (\ref{constraint}) and other linearized equations of motion,  then all other metric ansatz relating to it via ten dimensional diffeomorphism transformation would yield identical moduli space metric under the dimension reduction.
However to identify the precise metric ansatz corresponding to the fluctuation of a given modulus, additional information such as the preservation of certain global symmetries, is generally required.

\section{Coupling to D3 branes}\label{CouplingtoD3}
\paragraph{}
In this section we would like to consider coupling a D3 brane to the warped closed string background given by metric (\ref{WarpMetric}). Our aim here is to derive the K\"ahler potential for the position modulus of a mobile D3 brane.
In particular we will consider the simplified situation where only the unversal K\"ahler modulus $c(x)$ (and its imaginary partner) and the D3 brane position moduli $Y^i(x)$ are fluctuating. This is justified since a D3 brane does not source other moduli such as the dilaton $\tau$ or $C_2$.   
Earlier, the Hamiltonian provides a natural inner product on the space of metric fluctuations, where the orthogonality constraint to the unphysical diffeomorphism transformation can be imposed by the compensators. The question here would therefore be: What is the corresponding inner product on the vector space spanned by both closed and open string fluctuations, and the associated orthogonality constraints? 

Let us clarify the steps being taken here. We start with the dynamical warped background given by (\ref{WarpMetric}), and introduce a single spacetime filling D3 brane at a point $Y$ on the compact six manifold. 
The bosonic fluctuations of the D3 brane in such a background are encoded in the DBI+CS action $S_{\rm D3}$ 
\begin{equation}
S_{\rm D3}=S_{\rm DBI}+S_{\rm CS}=-T_3\int_{\cW_4} d^4\xi \sqrt{-\det({\rm{P}}[g])}+T_3\int_{\cW_4}{\rm P}\left[C_{(4)} \right]\,.\label{DBICSaction}
\end{equation}
Here ${\rm P}[g]$ and ${\rm P}\left[C_{(4)}\right]$ are the pullbacks of the bulk metric $g_{MN}$ and RR four form $C_{(4)}$
into the D3 brane world volume $\cW_4$. Here for our purpose of deriving the K\"ahler potential, we will ignore the worldvolume gauge fields and the pullback of the NS-NS two form, as their effects will mainly be modifying the partial derivatives into covariant ones \cite{Grana}.  
To incorporate (\ref{DBICSaction}) into the full ten dimensional action (\ref{IIBaction}), the localized action is then given by 
\begin{equation}
S_{\rm loc.}=2\kappa_{10}^2\int d^6y \delta^{(6)}(y-Y) S_{\rm D3}[Y]\label{Sloc}\,.
\end{equation}
Here we will present a complete derivation for the universal K\"ahler potential, and allow for both open and closed string degrees of freedom to fluctuate in the full action (\ref{IIBaction}). 
We will extract the relevant pieces in the D3 action (\ref{DBICSaction}), and then combine with the closed string action to obtain the full kinetic terms. In particular, due to the presence of the open string fluctuations, the orthogonality constraint (\ref{constraint}) will be modified. Although we will later focus on the case of the universal K\"ahler modulus, parts of our discussion will be applicable for other closed string moduli coupling to the open string degrees of freedom.

Let us focus on the kinetic terms of the scalar fields in the DBI Lagrangian (\ref{DBICSaction}), and the relevant terms are given by the pull-back of the metric (\ref{WarpMetric}) 
\begin{eqnarray}
{\rm P}(g)_{\mu\nu}&=&e^{2A(Y,u)+2\Omega(u)}\left\{\tilde{g}_{\mu\nu}(x)+2\partial_\mu\partial_\nu u^{I}(x)\bK_I(Y)+2\bB_{iI}(Y)\partial_\mu u^I(x) \partial_\nu Y^i\right\}\nonumber\\
&+&e^{-2A(Y,u)}\tilde{g}_{i{j}}(Y,u)\partial_\mu Y^i\partial_\nu {Y}^{{j}} \,.\label{PBmetric}
\end{eqnarray}
Here the indices $\mu\,,\nu$ run over the D3 brane worldvolume coordinates and we have made the static gauge choice $\xi^\mu=x^\mu$. 
Notice that the warped factor $e^{-4A(Y,u)}$ and the fiducial metric $\tg_{ij}(Y,u)$ are evaluated at the locus of the D3 brane in the compact Calabi-Yau manifold. 

The determinant of (\ref{PBmetric}) can be readily evaluated by using
\begin{equation}
\sqrt{\det({1+M})}=1+\frac{1}{2}{\rm{Tr}}(M)-\frac{1}{4}{\rm{Tr}}(M^2)+\frac{1}{8}({\rm{Tr}}(M))^2+\dots\,.
\end{equation}
which yields
{\small
\begin{equation}
-\frac{\cL^{\rm DBI}_{\rm kin}}{T_3\sqrt{-\tg_4}}=\frac{e^{2\Omega(u)}}{2}\tilde{g}_{ij}(Y,u)\partial_\mu Y^i\partial^\tmu Y^j+e^{4(A(Y,u)+\Omega(u))}(\bB_{iI}(Y)\partial_\mu u^I(x)\partial^\tmu Y^i)+\dots\label{LD3kin}
\end{equation}}
Here we have eliminated the vaccum energy $\propto e^{4(A(Y,u)+\Omega(u))}$ in (\ref{LD3kin}), as it will be cancelled by the pullback of $C_4$ in the CS term in a supersymmetric background specified by (\ref{backe4A}) and (\ref{ISD}) \cite{GKP}, and we have only kept terms up to quadratic order in the spacetime derivatives\footnote{To be clear, in factoring out the four dimensional spacetime metric from (\ref{PBmetric}), we have included $\partial_{\mu}\partial_{\nu}u^I\bK_I(y)$ term. Since we are only keeping quadratic fluctuations in the action, for most places, the contribution from $\partial_{\mu}\partial_{\nu} u^I \bK_I(y)$ is neglected. We come back to this point in section 4.}. 
The indice $\tmu$ here is to highlight that the summation in done using the unwarped four dimensional background metric $\tg_{\mu\nu}$.
The first term is the usual kinetic term for the transverse fluctuations $Y^i(x)$, whereas the others represent the non-trivial coupling between the closed string and the open string fluctuations. 
To complete the list, one would also need to include the fluctuations in the Chern-Simons term $S_{\rm CS}$, and couple them to the relevant RR kinetic terms in the supergavity action.  Instead of getting into that, however, one can already consider the possible modification to the orthogonality condition imposed by the metric compensators. This can be analysed from the Hamiltonian approach reviewed earlier, or from the more conventional Lagrangian approach used in \cite{STUD}. The equivalence between the two approaches relies on the fact that, in the presence of a D3 brane, for the metric ansatz of the form, the orthogonality conditions with the appropriately defined inner product remains the same as the linearized equations of motion.

To generalize the Hamiltonian approach to include D3 brane, we will again focus on the Einstein term in the SUGRA action and the kinetic terms given by the expansion of the DBI action (\ref{LD3kin}). 
Furthermore, we will assume for the moment, the RR fluctuations are freezed out.
The natural inner product can then be constructed from considering the following integral 
\begin{equation}
\bH^{\rm All}=\int d^9 x\left\{\dot{h}_{MN}\pi^{MN}-\cL^{G}_{\rm kin}+2\kappa^2_{10}\delta^{(6)}(y-Y)\cH^{\rm DBI}\right\}\,,\label{DefbH}
\end{equation}
which is a natural generalization of the closed string Hamiltonian (\ref{HGkin}).
Here the canonical momentum $\pi^{MN}$ is as defined in (\ref{PiMN}), whereas $\cL^G_{\rm kin}$ and $\cL_{\rm kin}^{\rm DBI}$ are given by (\ref{DefLG}) and (\ref{LD3kin}). We have also restored the six dimensional delta function $\delta^{(6)}(y-Y)$ and defined the Hamiltonian for the D3 brane $\cH^{DBI}$:
\begin{eqnarray}
\cH^{DBI}&=&g_{ij}(P^j \dot{Y}^i)-\cL^{\rm DBI}_{\rm kin}\nonumber\\
&=&-\eta_i P^i+\frac{1}{2}\frac{g_{ij}g_{tt}}{T_3\sqrt{-g_4}}P^iP^j+(\partial_\alpha Y^i~{\rm dependent~terms})\,,
\label{DefHDBI}
\end{eqnarray}
where  $\partial_\alpha Y^i$ denotes the partial derivative with respect to the three external spatial coordinates\footnote{Notice here that in the derivation of the Hamiltonian (\ref{DefHDBI}), we have neglected the terms proportional to $(\partial_{\mu} u^I B_{iI}(y))^2$, in the same approximation made in the Hamiltonian formulation of the closed string moduli.}. 
The canonical momentum $P^i$ is given by:
\begin{equation}
P_i=\frac{\partial\cL^{\rm DBI}_{\rm kin}}{\partial \dot{Y}^i}=T_3\sqrt{-g_4}(g^{tt}(\dot{Y}_i+\eta_i))=T_3\sqrt{-g_4} g^{tt} K_i\,,\label{DefPi}
\end{equation}
note that we have taken the convention $g_{tt}=-g_{00}>0$.
Similar to the extrinsic curvature $K^{MN}$ which is the physical metric fluctuations orthogonal to the spacelike hypersurface $\Sigma_t$, the vector $K^i$ has the natural geometric interpretation as the vector fluctuation orthogonal to $\Sigma_t$, and transforms under the nine-dimensional diffeomorphism of $\Sigma_t$.
In (\ref{DefbH}), the compensators $\eta_\alpha$ and $\eta_i$ defined in (\ref{DefCompensators}) again appear as non-dynamical Lagrangian multipliers. Due to the presence of D3 transverse fluctuations, however, the constraints imposed by $\eta_\alpha$ and $\eta_i$ now become:
\begin{eqnarray}
&&D^{M}\left(h^{-1/2}\pi_{M\alpha}\right)=0\label{SpacetimeConstraint}\,,\\
&&D^M(h^{-1/2}\pi_{Mi})+ \kappa^2_{10}{\delta^{(6)}(y-Y)}\frac{P_i}{\sqrt{h}}=0\,. \label{ModifiedConstraint}
\end{eqnarray}
Notice that we have assumed that $\tilde{g}_{\alpha 0}=\tg_{0\alpha}=0$, as the external spacetime is assumed to be maximally symmetric, so the external constraint (\ref{SpacetimeConstraint}) does not receive $\delta$-function corrections\footnote{In fact, to ensure that this equation is satisfied globally over a compact space, one also needs to include as usual a background charge density on the RHS of (\ref{ModifiedConstraint}).}. 
One can also verify that (\ref{SpacetimeConstraint}) and (\ref{ModifiedConstraint}) are equivalent to the linearized equations of motion in the presences of D3 branes:
\begin{equation}
\delta G_{0M}=\kappa_{10}^2\delta T^{(D3)}_{0M}\,,~~~M=1\,,\dots\,, 9\,.\label{LEOMD3}
\end{equation}

To see that $\bH^{\rm All}$ (\ref{DefbH}) with the constraint (\ref{ModifiedConstraint}) imposed is indeed the correct inner product, let us first write out the remainder of (\ref{DefbH}) after integrating out $\eta_N$:
\begin{equation}
\bH^{\rm All}
=\int d^9x \sqrt{h}\left[(g_{tt})^{1/2}h^{-1/2}\pi_{MN}K^{MN}+\kappa^2_{10}\frac{\delta^{(6)}(y-Y)}{\sqrt{h}} K^i P_i\right]\,.\label{DefbHIJ}
\end{equation} 
Here we have again taken the simpler case where $\partial_\mu u^{I}(x)=\delta_\mu^0\dot{u}^I$ and $\partial_\mu Y^i=\delta_\mu^0 \dot{Y}^i$ for extracting the kinetic energy.  
Now we can consider a nine-dimensional diffeomorphism transformation of $\Sigma_t$ acting on $K^{MN}$ and $K^i$, as generated by vector $V^{N}$, such that $K^{MN}$ transforms as in (\ref{Ktrans}) whereas:
\begin{eqnarray}
&&K^i~\longrightarrow~K^i+V^{i}\,,
\label{Trans2}
\end{eqnarray}
Substituting (\ref{Ktrans}) and (\ref{Trans2}) into (\ref{DefbHIJ}), one can show that, up to a total derivative term, the condition for the inner product to vanish on the unphysical diffeomorphism transformation is given by:
\begin{equation}
\int d^9 x\sqrt{h}\left\{V^{\alpha} D^M\left(h^{-1/2}\pi_{M\alpha}\right)+V^{i}
\left[D^{M}\left(h^{-1/2}\pi_{Mi}\right)+\kappa^2_{10}\delta^{(6)}(y-Y)\frac{P_i}{\sqrt{h}} \right]\right\}\label{difftrans2}
\end{equation}
Since $V^{N}$ is an arbitrary vector, the constraint (\ref{ModifiedConstraint}) in the compact directions combines with the one in the spacetime directions (\ref{SpacetimeConstraint}) ensure that the orthogonality conditions under diffeomorphism transformation are indeed imposed. Once again, due to such invariance of $\bf{H}^{\rm All}$ (\ref{DefbHIJ}), we can evaluate the moduli space metric in the convenient gauge, up to a reparametrization of the time variable. 
\paragraph{}
Having discussed how to couple a D3 brane to the universal K\"ahler modulus in the context of the Hamiltonian formalism, for the explicit computations of the moduli space metric, we will utilitize the diffeomorphism transformation on the metric (\ref{Ktrans}) and the vector (\ref{Trans2}) to work in the K-gauge given in (\ref{WarpMetricK}).    
This gauge holds distinct advantage that there are no cross couplings between terms such as $\partial_\mu u^I$ and $\partial_\mu Y^i$ as we can see in \eqref{LD3kin}.
Here in our analysis we will mostly work in the probe limit and neglect the change to the background metric caused by the motion of the mobile D3, and hence $Y^i$ is not a fluctuating modulus of the metric. 
We can then read off the kinetic energies for the universal K\"ahler modulus $c$ from the 10D Einstein term and for the D3 brane modulus from the DBI action. 
We will also comment on how  backreaction of the D3 brane can be incorporated in Hamiltonian formalism at the end of the section.     

Explicitly the metric fluctuations $\delta_I h_{MN}$ in the K-gauge (\ref{WarpMetricK}) take the following form:
\begin{eqnarray}
&&\delta_I h_{\alpha\beta}(x,u)=2\left(\frac{\partial A(y,u)}{\partial u^I}+\frac{\partial \Omega(u)}{\partial u^I}\right)e^{2A(y,u)+2\Omega(u)}\tilde{h}_{\alpha\beta}(x)\,,\label{deltahmunu}\\
&&\delta_I h_{ij}(y,u)=-e^{-2A(y,u)}\left(2\frac{\partial A(y,u)}{\partial u^I}\tilde{g}_{ij}(y,u)-\frac{\partial \tilde{g}_{ij}(y,u)}{\partial u^I}\right)\,.\label{deltahij}
\end{eqnarray}
where $\tilde{h}_{\alpha\beta}(x)=\tilde{g}_{\mu\nu}(x)$ for $\mu,\nu=1,2,3$. 
Here for the simplicity of notation, we have omitted the subscript $`` {\rm K}$" in the various quantities in (\ref{WarpMetricK}).
The canonical momenta $\delta_I \pi_{MN}$ are then given by:
\begin{eqnarray}
\delta_I \pi_{\alpha\beta}&=&\left(8\frac{\partial A(y,u)}{\partial u^I}-4\frac{\partial \Omega(u)}{\partial u^I}-\tg^{kl}(y,u)\frac{\partial \tg_{kl}(y,u)}{\partial u^I}\right)h_{\alpha\beta}(x,u)\,,\label{delpimunu}\\
\delta_I \pi_{ij}&=&\left(4\frac{\partial A(y,u)}{\partial u^I}-6\frac{\partial \Omega(u)}{\partial u^I}-\tg^{kl}(y,u)\frac{\partial \tg_{kl}(y,u)}{\partial u^I}\right)g_{ij}(y,u)+e^{-2A(y,u)}\frac{\partial \tg_{ij}(y,u)}{\partial u^I}\,,\label{delpiij}\\
P_i&=&T_3\sqrt{-g_4} g^{tt}\dot{Y}_i\label{Pi}\,.
\end{eqnarray}
Let us consider the constraint equations (\ref{SpacetimeConstraint}) and (\ref{ModifiedConstraint}), in the probe limit we will neglect the perturbation due to D3 on the warped metric, hence the $\delta^{(6)}(y-Y)$ term in (\ref{ModifiedConstraint}). One can next show that the constraint (\ref{SpacetimeConstraint}) can be trivially satisfied, while (\ref{ModifiedConstraint}) yields:
\begin{eqnarray}
&&D^M((g^{tt})^{1/2}\delta_I \pi_{Mi})\nonumber\\
&&=-(g^{tt})^{1/2}\left(e^{4A}\partial_i \partial_I e^{-4A(y,u)}+\tilde{\nabla}^j(\partial_I \tg_{ij}-\tg_{ij}(\tg^{kl}\partial_I\tg_{kl}))+4\partial^{\tilde{j}}A \partial_I \tg_{ij}-2\partial_iA\tg^{kl}\partial_I\tg_{kl}\right)=0\,.\nonumber\\
\label{Explicitconstraint2}
\end{eqnarray}  
Now for the case of universal K\"ahler modulus $u^I=c$, a viable solution to (\ref{Explicitconstraint2}) is then given by:
\begin{equation}
\partial_c\tg_{ij}=0\,,~~~ e^{-4A(y,c)}=e^{-4A_0(y)}+c\,.\label{defe4A}
\end{equation}
Notice that one can in fact replace $c$ by an arbitrary function $f(c)$, but this will become a mere field redefinition of the universal K\"ahler modulus. 
This solution (\ref{defe4A}) fits well with the usual physical interpretation that the universal K\"ahler modulus $c$ corresponds to an overall rescaling of the internal space and preserve all of its isometries, and hence unwarped metric $\tg_{ij}$ should not have explicit dependence on $c$. Furthermore $c$ remains massless even in the presence of fluxes, which can only be achieved if $c$ (or more generally $f(c)$) appears as an undetermined integration constant in the defining equation of the warp factor.

At this point, we emphasize that the solution \eqref{defe4A} is by no means unique. At least, we could obtain a family of solutions generated by the gauge transformation $\zeta_i = ce^{2A+2\Omega} \partial_i\Lambda$, $\zeta_\mu = -\partial_\mu c e^{2A+2\Omega} \Lambda$ that preserves the K-gauge condition. We regard $\partial_c\tg_{ij}=0$ as a gauge condition to define our universal K\"ahler moduli in a conventional way. In other words, the universal K\"ahler moduli is defined as an overall scaling of the metric only in this specific K-gauge. With a different gauge choice, the independence of the internal metric $\tilde{g}_{ij}$ on $c$ no longer holds in general. 

It is useful to note that our background metric with no D3-brane is diffeomorphic to the one presented in \cite{FTUD}. To see this, we can perform the gauge transformation $\zeta_i = 0$, $\zeta_\mu = ce^{2A+2\Omega} \partial_i \mathbf{K}$. As we discussed before, such a diffeomorphism does not change $\tilde{g}_{ij}$ and $e^{4A}$. As a consequence, the constraint equations derived there should be equivalent to ours with no change of variables albeit it is much more tedious to obtain them directly in their gauge.\footnote{We would like to thank A.~Frey for helpful communication on this point.} Once we introduce the D3-brane, however, the gauge transformation acts on the DBI action non-trivially, and the advantage to use the K-gauge becomes distinctive.

Using (\ref{defe4A}), which implies that the internal unwarped metric is independent of the universal K\"ahler modulus $c$, 
we can write down an explicit expression for the Weyl factor $e^{-2\Omega}$ from the definition (\ref{e2Omega}):
\begin{equation}
e^{-2\Omega(c)}=c+\frac{V_W^{0}}{V_{CY}}\,.\label{explicite2Omega}
\end{equation}
where $V_W^{0}=\int d^6 y\sqrt{\tg_6}e^{-4A_0(y)}$. 
Combining (\ref{defe4A}) and (\ref{explicite2Omega}), we can now write down the kinetic terms for both the universal K\"ahler modulus $c$ and the D3 brane modulus $Y^i$ using (\ref{DefbHIJ}):
\begin{equation}
S_{\rm kin.}^{\rm All}=\frac{1}{2\kappa^{2}_{10}}\int dt \bH^{\rm All}
=\frac{3}{\kappa_4^2}\int d^4 x\sqrt{-\tg_4}\left(\frac{1}{(2e^{-2\Omega})^2}\tg^{tt}(\dot{c})^2
+\frac{T_3\kappa_4^2}{3}\frac{1}{(2e^{-2\Omega})}\tg^{tt}\tg_{ij}\dot{Y}^i\dot{Y}^j\right)\,,\label{SkinAll}
\end{equation}  
where we have used the relation $\kappa_{10}^2=\kappa_4^2 V_{CY}$.
In contrast with \cite{DT,FTUD}, here in deriving (\ref{SkinAll}), it is crucial to keep the $\delta_c \pi_{\alpha\beta}$ contribution to kinetic energy, as we do not see obvious reasons for setting it to zero from the constraint equations. 
So far in the K-gauge we used (\ref{WarpMetricK}), the explicit compensator dependence $\propto \ddot{c}\bK_c$ only appears in $g^{tt}$ component. Its explicit form is not determined through the constraint equations and does not enter the kinetic terms at the order of quadratic fluctuations in the absence of the potential.
However even for $c\equiv c(t)$, it is crucial to have a non-vanishing $\bK_c(y)$ to satisfy the $(1,1)$ components of the linearized Einstein equation:
\begin{equation}
\tilde{\nabla}^2 \bK_c(y)=e^{-4A_0(y)}-\frac{V_{W}^{0}}{V_{CY}}\,.\label{DefeqnKc}
\end{equation} 
Here we have used the explicit expression for $e^{-4A(c)}$ (\ref{defe4A}) and $e^{-2\Omega(c)}$ (\ref{explicite2Omega}) to simplify the expression. Similar equations were also noticed in \cite{GM, FTUD}. As will be discussed later, 
for moduli which develop a potential (e.g., complex structure moduli in flux compactification),
a non-trivial $\bK_c(y)$ can give rise to an additional contribution to the kinetic term for such moduli.

From (\ref{SkinAll}), we can finally write down the K\"ahler potential:
\begin{equation}
\kappa_4^2\cK(\rho,Y)=-3\log\left[\rho+\bar{\rho}-\gamma k(Y,\overline{Y})+2\frac{V_W^0}{V_{CY}}\right]\,,~~~\gamma=\frac{T_3\kappa_4^2}{3}\,,\label{D3KahlerPot}
\end{equation}
where $V_{W}^0$ denotes the warped volume in the absence of the D3 brane and and $k(Y,\overline{Y})$ is the little K\"ahler potential for the unwarped internal manifold. The holomorphic volume modulus $\rho$ is defined to be:
\begin{equation}
\rho=\left(c+\frac{\gamma}{2}k(Y,\overline{Y})\right)+i\chi\,.\label{defrho}
\end{equation} 
where the axionic partner $\chi$ of $c$ comes from dimension reduction of $C_4$. Our definition of $\rho$ was motivated by a non-trivial $U(1)$ fibration of $\chi$ over the $Y$-moduli space. This arises from the pullback of $C_4$ onto the D3 brane world volume and manifest itself at weak warping in following transformation \cite{KKLMMT}\footnote{We are grateful to Juan Maldacena for discussing this point with us.}:
\begin{equation}
\rho\rightarrow \rho+ \gamma f(Y)\,,~~\bar{\rho}\rightarrow \bar{\rho}+\gamma\overline{{f}(Y)}\,,~~k(Y,\overline{Y})\rightarrow k(Y,\overline{Y})+ f(Y)+\overline{{f}(Y)}\,.\label{U1trans}
\end{equation}
This is to ensure that the overall K\"ahler potential is invariant under the 
little K\"ahler transformation of $k(Y,\overline{Y})$.
The K\"ahler potential (\ref{D3KahlerPot}) we derived at strong warping from direct dimension reduction remains consistent with such a fibration, 
and also allows us to determine the precise value of $\gamma$ and other moduli-independent 
constants such as $V^0_W/V_{CY}$. 
In the absence of a D3 brane, we 
obtain the K\"ahler potential for the universal K\"ahler modulus which takes the same form presented in \cite{FTUD} though we derive it here consistently in a single gauge.
\paragraph{}
Let us finish the section on D3 brane by commenting on the issue of the ``rho problem" and discussing how it can be resolved by including the backreaction of the D3-brane on the warp factor. In the probe limit we have discussed so far, the definition of the holomorphic K\"ahler modulus $\rho$ yields:
\begin{equation}
c=\frac{1}{2}(\rho+\bar{\rho}-\gamma k(Y,\bar{Y}))\,.\label{Nobackc}
\end{equation}
If one considers a spacetime-filling D7 brane wrapping on some supersymmetric four cycle $\cC_4$ in the warped background (\ref{WarpMetric}), one can show that the gauge kinetic function is proportional to the following integral in the strongly warped limit \cite{MMS}:
\begin{equation}
g_7^{-2}\propto \int_{\cC_4} d^4 y \sqrt{\hat{G}_4} e^{-4A(y,c)}=\int_{\cC_4} d^4 y \sqrt{\hat{G}_4}(c+ e^{-4A_0(y)})=c V_4+V_4^{W} \label{Defg7}\,.
\end{equation}
Here $\hat{G}_4$ denotes the pullback of the unwarped metric $\tg_6$ onto $\cC_4$, and in the second equality we have used the solution (\ref{defe4A}) and  $V_4^W$ denotes the moduli independent part of the warped 4-cycle volume.
Comparing (\ref{Nobackc}) and (\ref{Defg7}), we see, due to the little K\"ahler potential $k(Y,\overline{Y})$, that $g_7^2$ is not the real part of a holomorphic function on the brane moduli space. Supersymmetry requires the D7 gauge kinetic function to be a holomorphic function of moduli and so as in the weakly warped situation \cite{DG,Baumann1}, there is a rho problem.

As also pointed out in \cite{Baumann1}, the resolution to such rho problem is to properly include the backreaction of the D3 brane on the warp factor. In the context of the Hamiltonian formalism we discussed above, this translates into the inclusion of the $\delta$-function term in the constraint equation (\ref{ModifiedConstraint}). 
In other words, the D3 brane position $Y^i$ now also becomes a modulus of the bulk metric. The correction to the warp factor $e^{-4A}$ can be determined through the linearized equation of motion for the RR four-form, perturbed by localized source \cite{GM,Baumann1}. 
The resultant warp factor now becomes:
\begin{equation}
e^{-4A(y,c,Y)}=c+e^{-4A_0(y)}+\frac{\gamma}{2} k(Y,\overline{Y})+{[\rm hol.~+~antihol.]}\label{e4AYc}\,.
\end{equation}
Here the additional holomorphic and anti-holomorphic pieces satisfy the Laplace
equation and contains pieces that generate the $\delta(y-Y)$, and also relevant
to make $k(Y,\bar{Y})$ Kahler invariant. With this understanding, one can show
that (\ref{e4AYc}) is consistent with the modified constraint
(\ref{ModifiedConstraint}).
Upon substituting (\ref{e4AYc}) instead of (\ref{defe4A}) into (\ref{Defg7}), one can now see that the integral becomes:
\begin{eqnarray}
&&\int_{\cC_4} d^4 y \sqrt{\hat{G}_4} e^{-4A(y,c,Y)}= V_4\left(c+\frac{\gamma}{2} k(Y,\overline{Y})\right)+V_4^{W}+{[\rm hol.~+~antihol.]}\nonumber\\
&&=\frac{1}{2}(\rho+\overline{\rho}) V_4  + V_4^W + {[\rm hol.~+~antihol.]}\,.\label{newg7}
\end{eqnarray}
In other words the D7 gauge coupling is now the real part of a holomorphic function, as demanded by supersymmetry. We therefore resolved the rho problem in the general warped background.
This argument should be consistent with the field redefinition made in (\ref{Nobackc}). It suggests the effective warp factor in the closed string sector after taking into account the backreaction of 
open strings must be accompanied by the shift $\frac{\gamma}{2}k(Y,\bar{Y})$ as in eq.~(\ref{e4AYc})\footnote{Note that one  cannot substitute this backreacted warp factor into the DBI action of the D3-brane due to the well known
self energy problem.}.

\section{Discussions}\label{Discussions}
\paragraph{}
In this paper, we developed a general approach to compute K\"ahler potentials involving simultaneously open and closed string moduli in warped compactification. 
As an illustration, we
 considered 
 the position moduli of a D3-brane coupled to the universal K\"ahler modulus of a warped background
 and computed the combined K\"ahler potential. 
By restricting our results to only the closed string fluctuations, we also clarified some previous closed string derivations in the literature, in particular, the subtleties with gauges.

There are several potential applications of this work, one of which is a precise determination of the inflaton potential for D-brane inflation 
 \cite{stringcosmologyreviews, Baumann:2008kq, DelicateU, OpenRacetrack}. 
As is well known, if the inflaton potential is generated by an F-term in supergravity, inflationary physics which is determined by the slow-roll parameter $\eta$ is sensitive even to dimension six Planck suppressed operators.
Therefore, the warped K\"ahler potential for the D3-brane moduli we obtained
can play a crucial role in determining the eta parameter for this broad class of models. The combined K\"ahler potential is also an essential piece of information for determining the D3-brane vacua \cite{DeWolfe:2007hd,Brown:2008zq} 
in strongly warped backgrounds.

We should point out that although the convenient gauge choice we suggested is useful and simple for extracting the kinetic terms for moduli that do not develop a classical potential, complications can arise for fields that appear into the classical potential. Examples include complex structure moduli in flux compactification.
This is because the compensator
$K \ddot{u} dt^2$ contributes to $\sqrt{g_{tt}}$ which multiplies the potential in the action. Upon integrating $\int d^{10} x\sqrt{-g_{10}}~V(u)$ by parts, the compensator field $K$ gives rise to an additional contribution $\sim K V'(u) \dot{u}^2$ to the kinetic terms. In fact, a naive extrapolation of our approach to the complex structure modulus of the conifold without taking this subtlety into account leads to a K\"ahler metric which is not positive definite, though on dimensional ground, one would expect the same parametric dependence as the expressions presented in \cite{DST,DT}.

Another subtlety is that the potential term explicitly breaks the assumption
that (arbitrary) constant $u$ solves the background (zero-th) order equations of
motion. Thus, there is less sense to talk about the linearlized equations of
motion around the background if we are away from the vacua. To determine the
non-supersymmetric vacua, we typically need to compute the K\"ahler potential.
The linearized equations of motion and the Hamiltonian formulation, however,
require that we should expand around the (yet-to-be-determined) vacua. It
appears to be a chicken and egg problem, and approximate iterative or bootstrap
approach might be necessary. The applicability of the Hamiltonian formulation to
obtain kinetic terms away from the vacua remains open.

It would be interesting to derive explicitly K\"ahler potentials for moduli of this type. We leave this and related problems for future work.

\subsection*{Acknowledgements}
\paragraph{}
We thank Andrew Frey, Juan Maldacena, Fernando Marchesano, Paul McGuirk,
Nathan Seiberg, Gonzalo Torroba, and Bret Underwood  for comments and discussions. HYC would also like to specially thank Peter Ouyang for various insightful comments.
We are grateful to Andrew Frey, Peter Ouyang, Fernando Quevedo, and Bret Underwood for comments on this manuscript. 
The work of HYC and GS  was supported in part by NSF CAREER Award No. PHY-0348093, DOE grant DE-FG-02-95ER40896, a Research Innovation Award and a Cottrell Scholar Award from Research
Corporation, a Vilas Associate Award from the University of Wisconsin, and a John Simon Guggenheim Memorial Foundation Fellowship. GS would also like to acknowledge support from the Ambrose Monell Foundation during his stay at the Institute for Advanced Study.
The work of YN
was supported in part by the National Science Foundation under Grant
No. PHY05-55662 and the UC Berkeley Center for Theoretical Physics.
HYC and GS thank the Institute for Advanced Study and Physics Department, Princeton University
for hospitality and support while this work was written.

\end{document}